\documentclass[Physsubmission, Phys]{SciPost}
\hypersetup{
    colorlinks,
    linkcolor={red!50!black},
    citecolor={blue!50!black},
    urlcolor={blue!80!black}
}

\usepackage[bitstream-charter]{mathdesign}
\usepackage{graphics}
\graphicspath{{./figures/}}

\newcommand{\tcite}[1]{~\cite{#1}}
\newcommand{\tref}[1]{~\ref{#1}}

\DeclareSymbolFont{usualmathcal}{OMS}{cmsy}{m}{n}
\DeclareSymbolFontAlphabet{\mathcal}{usualmathcal}

\begin{document}

% TODO: write your article's title here.
% The article title is centered, Large boldface, and should fit in two lines
\begin{center}{\Large \textbf{
A spectator-model way to transverse-momentum-dependent gluon distribution functions
}}\end{center}

% TODO: write the author list here. Use initials + surname format.
% Separate subsequent authors by a comma, omit comma at the end of the list.
% Mark the corresponding author with a superscript *.
\begin{center}
Alessandro Bacchetta\textsuperscript{1,2},
Francesco Giovanni Celiberto\textsuperscript{3,4,5$\star$},
Marco Radici\textsuperscript{2} and
Pieter Taels\textsuperscript{6}
\end{center}

% TODO: write all affiliations here.
% Format: institute, city, country
\begin{center}
{\bf 1} Dipartimento di Fisica, Universit\`a di Pavia, via Bassi 6, I-27100 Pavia, Italy
\\
{\bf 2} INFN Sezione di Pavia, via Bassi 6, I-27100 Pavia, Italy
\\
{\bf 3} European Centre for Theoretical Studies in Nuclear Physics and Related Areas (ECT*), I-38123 Villazzano, Trento, Italy
\\
{\bf 4} Fondazione Bruno Kessler (FBK), I-38123 Povo, Trento, Italy
\\
{\bf 5} INFN-TIFPA Trento Institute of Fundamental Physics and Applications, \\ I-38123 Povo, Trento, Italy
\\
{\bf 6} Centre de Physique Th\'eorique, \'Ecole polytechnique, CNRS, \\ I.P. Paris, F-91128 Palaiseau, France
\\
% TODO: provide email address of corresponding author
* fceliberto@ectstar.eu
\end{center}

\begin{center}
\today
\end{center}

% For convenience during refereeing (optional),
% you can turn on line numbers by uncommenting the next line:
%\linenumbers
% You should run LaTeX twice in order for the line numbers to appear.

\definecolor{palegray}{gray}{0.95}
\begin{center}
\colorbox{palegray}{
  \begin{tabular}{rr}
  \begin{minipage}{0.1\textwidth}
    \includegraphics[width=22mm]{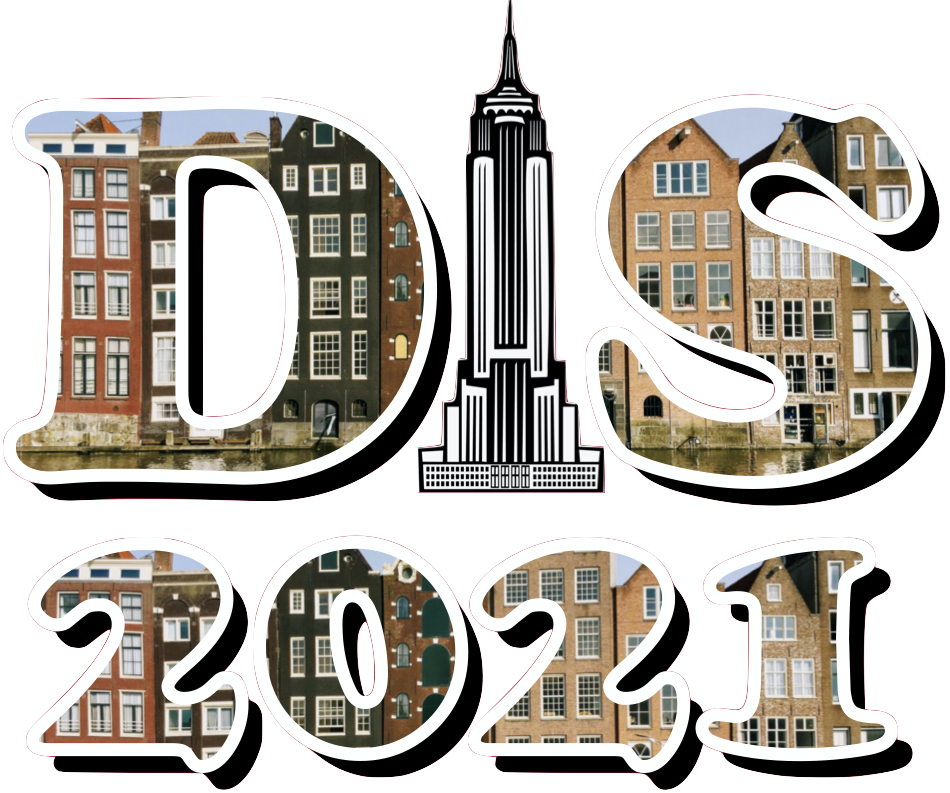}
  \end{minipage}
  &
  \begin{minipage}{0.75\textwidth}
    \begin{center}
    {\it Proceedings for the XXVIII International Workshop\\ on Deep-Inelastic Scattering and
Related Subjects,}\\
    {\it Stony Brook University, New York, USA, 12-16 April 2021} \\
    \doi{10.21468/SciPostPhysProc.?}\\
    \end{center}
  \end{minipage}
\end{tabular}
}
\end{center}

\section*{Abstract}
{\bf
We present exploratory analyses of the 3D gluon content of the proton via a study of unpolarized and polarized gluon TMDs at twist-2, calculated in a spectator model for the parent nucleon. Our approach embodies a flexible parametrization for the spectator-mass function, suited to describe both moderate and small-$x$ effects. All these studies can serve as a useful guidance in the investigation of the gluon dynamics inside nucleons and nuclei, which constitutes one of the major goals of new-generation colliding machines, as the EIC, the HL-LHC, NICA, and the FPF.
}

\vspace{10pt}
\noindent\rule{\textwidth}{1pt}
\tableofcontents\thispagestyle{fancy}
\noindent\rule{\textwidth}{1pt}
\vspace{10pt}

\section{Introduction}
\label{sec:intro}

The study of the proton content via transverse-momentum-dependent (TMD) parton distribution functions represents a challenging line of research plans at current and new-generation colliding machines. While in the last years the investigation of the quark-TMD field has reached important milestones, from the deep knowledge of formal properties to the more and more accurate extraction of quark densities from global fits, the gluon-TMD sector still represents a largely unexplored territory.
A first classification of unpolarized and polarized gluon TMD distributions was first made in Ref.\tcite{Mulders:2000sh} and subsequently extended in Refs.\tcite{Meissner:2007rx,Lorce:2013pza,Boer:2016xqr}. Recent phenomenological analyses on gluon TMDs can be found in Refs.\tcite{DAlesio:2017rzj,DAlesio:2018rnv,Bacchetta:2018ivt,DAlesio:2019qpk,Boer:2020bbd}.
A major difficulty that emerges in formal studies of gluon TMDs is their process dependence. Different kinds of reactions are sensitive to distinct \emph{gauge-link} structures, and this leads to a more intricate \emph{modified universality} with respect to what we observe for quark TMDs. Two main gauge links can be identified. They have been classified in the context of small-$x$ analyses as Weisz\"acker--Williams and dipole TMDs~\cite{Dominguez:2011wm}. They are strictly related to gluon correlators where for $T$-odd TMDs the $f_{abc}$ and $d_{abc}$ QCD color structures respectively emerge. Therefore, they are also known among the TMD community as $f$-type and $d$-type gluon TMDs.
At low-$x$ values and large transverse momenta, the gluon content of the proton is described by the so-called \emph{unintegrated gluon distribution} (UGD), whose evolution is governed by the Balitsky--Fadin--Kuraev--Lipatov (BFKL) equation\tcite{Fadin:1975cb,Balitsky:1978ic} (for recent applications see Refs.~\cite{Bolognino:2018rhb,Celiberto:2019slj,Bolognino:2019pba,Brzeminski:2016lwh,Celiberto:2018muu,Celiberto:2020wpk,Bolognino:2019yls,Celiberto:2020tmb,Celiberto:2021dzy,Celiberto:2021fdp,Celiberto:2022dyf}). Its relation to the low-$x$ limit of gluon TMDs and, more in general, to the Collins--Soper--Sterman (CSS) evolution\tcite{Collins:2011zzd,Collins:1981uk} has been investigated in Refs.\tcite{Dominguez:2011wm} and\tcite{Nefedov:2021vvy,Hentschinski:2021lsh}, respectively.
In this work we present a study on leading-twist $T$-even gluon TMDs calculated in a \emph{spectator model} for the parent proton. Our framework is suited to analyses both in moderate and small-$x$ ranges.

\section{TMD gluon distribution functions}

According to the spectator-model approximation, the proton can emit a gluon with longitudinal-momentum fraction $x$ and transverse momentum $\boldsymbol{p}_T$, and the remainders are treated as an effective colored particle with mass $M_X$ and possessing the quantum numbers of a fermion, that we call spectator. The nucleon-gluon-spectator coupling is encoded in a effective vertex that contains two form factors, chosen as dipolar functions of $\boldsymbol{p}_T^2$. The main advantage of using dipolar form factors consists in the possibility of cancelling gluon-propagator singularities, quenching the effects of large transverse momenta where a pure TMD description is not anymore adequate, and removing logarithmic divergences emerging in $\boldsymbol{p}_T$-integrated densities.

In Ref.\tcite{Bacchetta:2008af} a pioneering study on quark TMDs was proposed, by considering different di-quark spectator polarization states and nucleon-parton-spectator form factors. In Ref.\tcite{Bacchetta:2010si} the weight of azimuthal asymmetries was assessed.

\begin{figure}[h]
\centering
\includegraphics[width=0.45\textwidth]{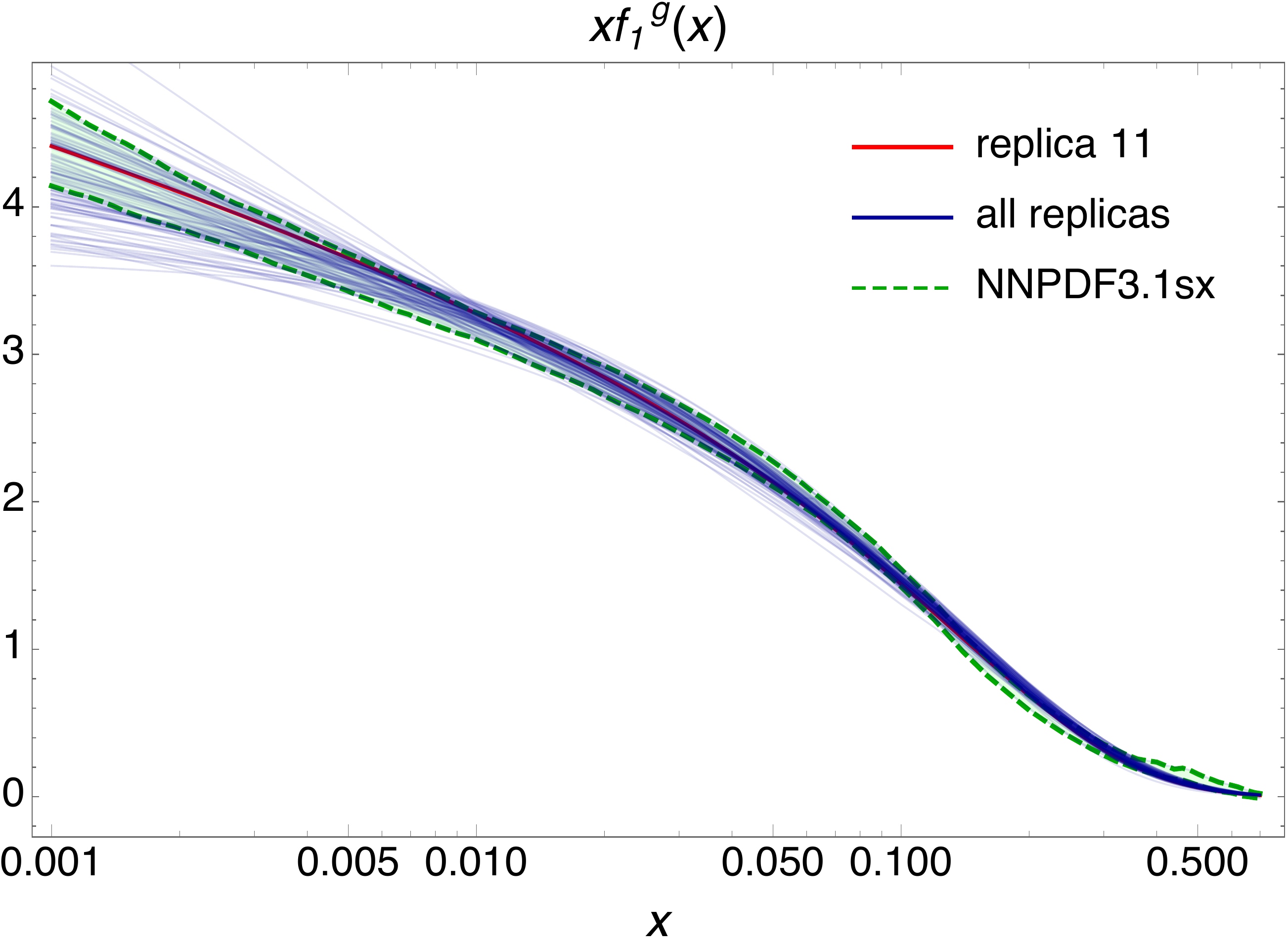} \hspace{0.5cm}
\includegraphics[width=0.4675\textwidth]{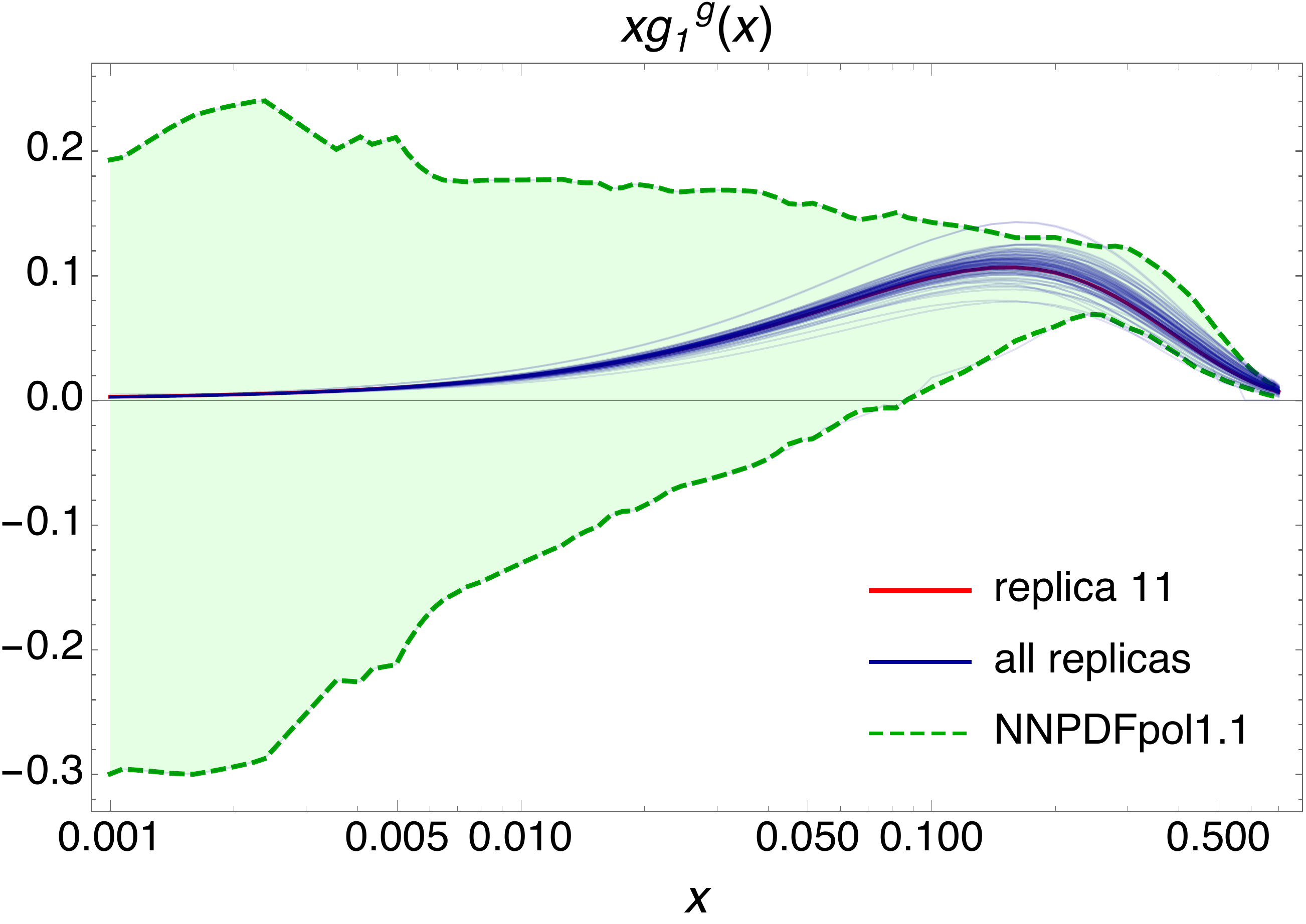}
\caption{$x$-dependence of the unpolarized (left) and helicity (right) gluon PDFs densities calculated in the spectator model at the initial scale $Q_0 = 1.64$ GeV. Green bands with dashed borders stand for the {\tt NNPDF3.1x}\tcite{Ball:2017otu} and the {\tt NNPDFpol1.1}\tcite{Nocera:2014gqa} parametrizations. Blue curves depict the 100 replicas for our integrated TMDs. Red curve for the most representative replica \#11.}
\label{fig:PDF_f1_g1}
\end{figure}

\begin{figure}[h]
\centering
\includegraphics[width=0.45\textwidth]{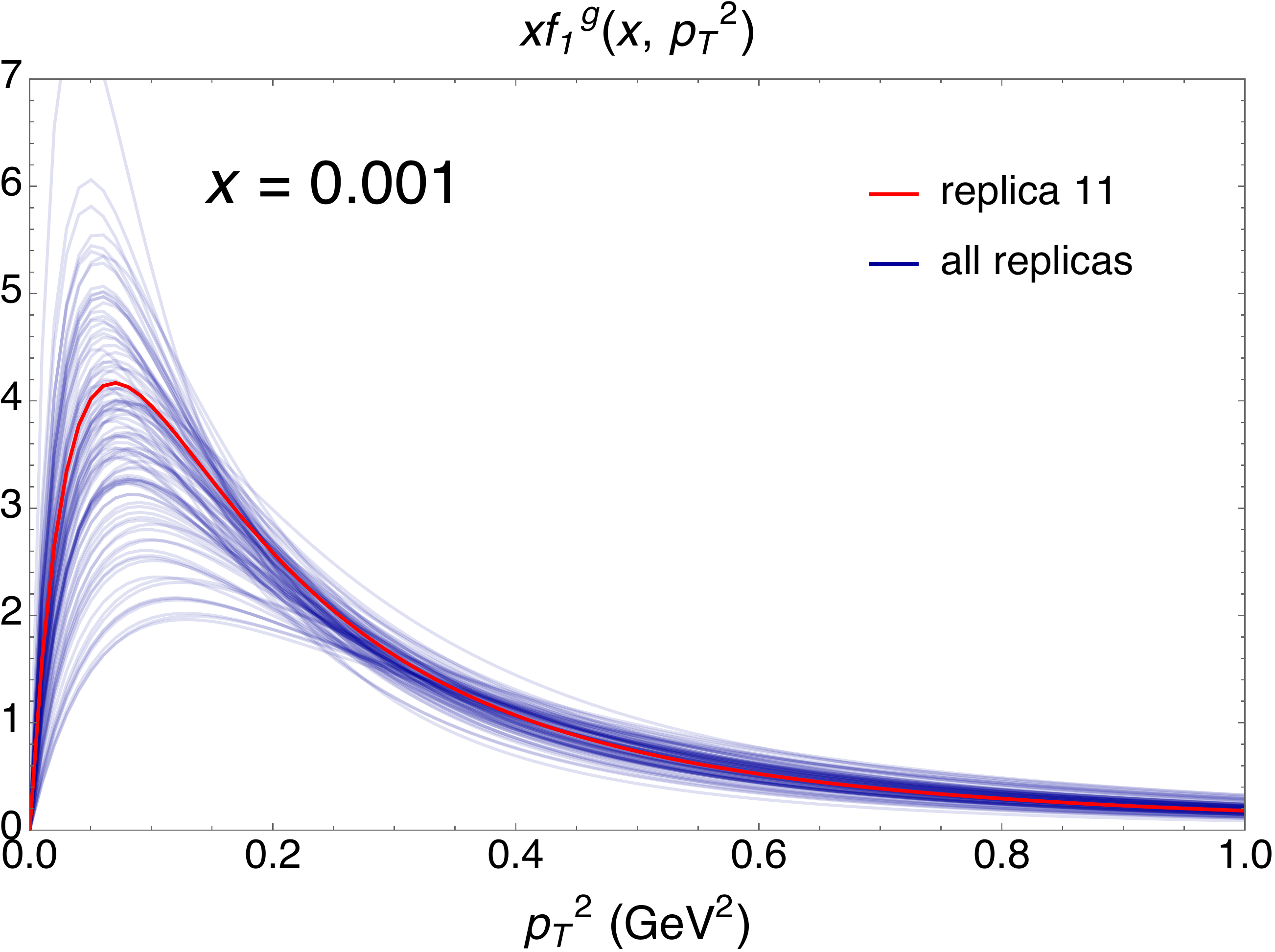} \hspace{0.5cm}
\includegraphics[width=0.465\textwidth]{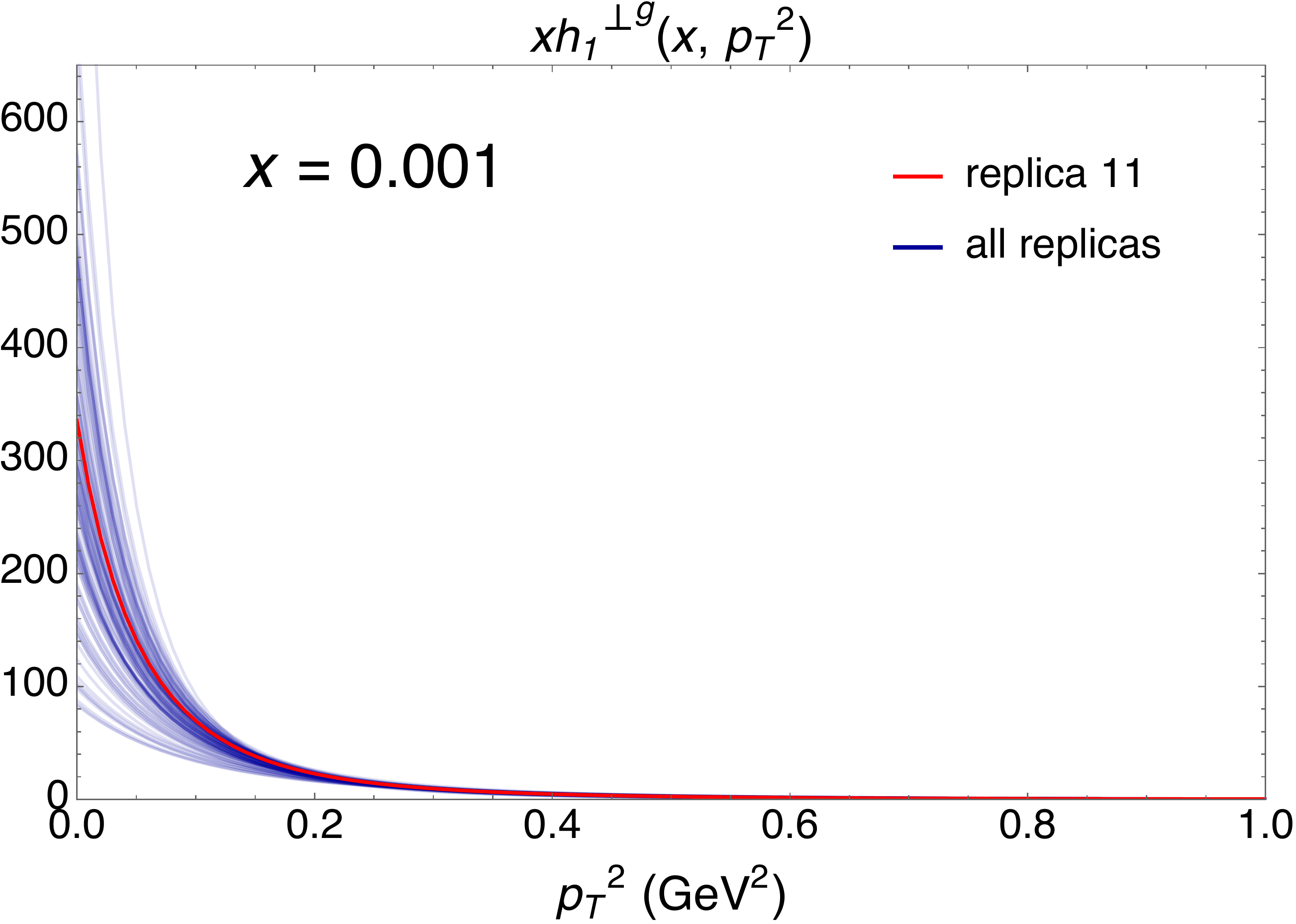}
\caption{Transverse-momentum dependence of the unpolarized (left) and Boer--Mulders (right) gluon TMD densities calculated in the spectator model, for $x=10^{-3}$ and at the initial scale $Q_0 = 1.64$ GeV. Red curve for the most representative replica \#11.}
\label{fig:TMD_f1_h1p}
\end{figure}

In the present study we present our calculation in the spectator model of $T$-even gluon TMDs at twist-2. We improved the genuine spectator-model approach by allowing the spectator mass, $M_X$, to be in a range of values weighed by the following 7-parameter spectral function
\begin{equation}
\label{eq:rho}
 \rho_{\rm [spect.]} (M_X) = \mu^{2a} \left( \frac{A}{B + \mu^{2b}} + \frac{C}{\pi \sigma} e^{-\frac{(M_X - D)^2}{\sigma^2}} \right) \;.
\end{equation}
The expression for a given TMD reads
\begin{equation}
\label{eq:TMD}
 {\cal F}^g(x,\boldsymbol{p}_T^2) = \int_M^{\infty} d M_X \, \rho_{\rm [spect.]} (M_X) \, \hat{\cal F}^g(x, \boldsymbol{p}_T^2; M_X) \;,
\end{equation}
with $\hat{\cal F}^g$ the corresponding TMD obtained in a pure spectator-model calculation.
Model parameters were fitted to simultaneously reproduce the gluon unpolarized ($f_1^g(x)$) and helicity ($g_1^g(x)$) collinear parton distribution functions (PDFs), obtained in global fits at the initial scale $Q_0 = 1.64$ GeV (see Fig.\tref{fig:PDF_f1_g1}). We performed our fit by making use of the so-called bootstrap method.
We created $N$ \emph{replicas} of the central value of the {\tt NNPDF} parametrization by randomly varying it with a Gaussian
noise that keeps the same variance of the original parametrization uncertainty. We fitted each replica separately and we
obtained $N$-dimensional vector for each parameter of the model. A complete description of our model together all technical details of our fit procedure can be found in Ref.~\cite{Bacchetta:2020vty} (see also Refs.\tcite{Celiberto:2021zww,Bacchetta:2021lvw,Bacchetta:2021twk,Bacchetta:2022esb,Bolognino:2022uty,Celiberto:2022fam}).
We show in Fig.\tref{fig:TMD_f1_h1p} the $\boldsymbol{p}_T^2$-dependence of two $T$-even gluon TMDs calculated at $x = 0.001$ and at the same initial scale, $Q_0 = 1.64$ GeV. Each one of our TMDs exhibit a distinct shape.
The unpolarized gluon density $x f_1^g(x, \boldsymbol{p}_T^2)$ (left panel) shows a non-Gaussian pattern in $\boldsymbol{p}_T^2$, a large flattening tail in the $\boldsymbol{p}_T^2 \to 1$ GeV limit, and it goes to a quite small value when $\boldsymbol{p}_T^2 \to 0$. Conversely, the Boer--Mulders gluon distribution $x h_1^{\perp g}(x, \boldsymbol{p}_T^2)$ (right panel), that is connected to the density of transversely polarized gluons inside an unpolarized proton, starts from a finite value at $\boldsymbol{p}_T^2 = 0$ and decreases very fast when $\boldsymbol{p}_T^2$ grows.

\section{Conclusion}

We presented a model dependent calculation of all twist-2 $T$-even gluon TMDs based on the assumption that what remains of the proton after gluon emission can be described as an effective spin-1/2 spectator particle.
We improved the genuine spectator-model description by weighing its mass via a versatile spectral function.
We fitted model parameters to reproduce the $x$-shape of collinear unpolarized and helicity gluon PDFs that were extracted from global fits.
At the current level, our model does not incorporate any gauge-link dependence, and the extension to twist-2 $T$-odd gluon TMD distributions is underway.
Another intriguing perspective is represented by encoding in the description of the unpolarized gluon TMD genuine small-$x$ effect from the BFKL resummation\tcite{Fadin:1975cb,Balitsky:1978ic}.
Exploratory studies on gluon-TMD phenomenology via our model can represent a useful guidance in accessing the proton content at new-generation colliding machines, as the \emph{Electron-Ion Collider}~(EIC)~\cite{AbdulKhalek:2021gbh}, the \emph{High-Luminosity Large Hadron Collider} (HL-LHC)~\cite{Chapon:2020heu}, NICA~\cite{Arbuzov:2020cqg}, and the \emph{Forward Physics Facility}~\cite{Anchordoqui:2021ghd,Feng:2022inv}.

\bibliography{references.bib}

\nolinenumbers

\end{document}